\documentclass[dcolumn,showpacs]{revtex4}
\usepackage[xdvi]{graphicx}
\usepackage{dcolumn}
\usepackage{amsmath}
\makeatletter
\def\btt#1{\texttt{\@backslashchar#1}}%
\DeclareRobustCommand\bblash{\btt{\@backslashchar}}%
\makeatother

\newcommand{\bra}{\left\langle}
\newcommand{\ket}{\right\rangle}
\newcommand{\con}{_{\rm con}}
\newcommand{\kin}{_{\rm kin}}

\begin{document}
\title{Measuring Nonequilibrium Temperature of Forced Oscillators}
\author{Takahiro Hatano}
\affiliation{Center for Promotion of Computational Science and Engineering, 
Japan Atomic Energy Research Institute, Ibaraki 319-1195, Japan}
\author{David Jou}
\affiliation{Departament de F\'\i sica, 
Universitat Aut{\` o}noma de Barcelona, 08193 Bellaterra, Catalonia, Spain}
\affiliation{Institut d'Estudis Catalans, Carme 47, 08001 Barcelona, 
Catalonia, Spain}
\date{\today}

\begin{abstract}
The meaning of temperature in nonequilibrium thermodynamics 
is considered by using a forced harmonic oscillator in a heat bath, 
where we have two effective temperatures 
for the position and the momentum, respectively. 
We invent a concrete model of a thermometer to testify 
the validity of these different temperatures 
from the operational point of view. 
It is found that the measured temperature depends on 
a specific form of interaction between the system and a thermometer, 
which means the zeroth law of thermodynamics 
cannot be immediately extended to nonequilibrium cases.
\end{abstract}

\maketitle

\section{INTRODUCTION}

Temperature and entropy are basic concepts of thermodynamics 
which have clear definitions and meaning in equilibrium 
but which are not yet fully understood in nonequilibrium situations.
In equilibrium thermodynamics, one way to introduce temperature is 
to define entropy somehow (e.g. through the adiabatic invariant) 
so that temperature can be introduced as a derivative of entropy 
with respect to energy;
\begin{equation}
\label{def1}
\beta =\frac{\partial S}{\partial U}.
\end{equation}
However, since nonequilibrium entropy has never been 
constructed in a consistent way,
we cannot define nonequilibrium temperature in this manner.
For example, we don't know whether the entropy is a measurable quantity 
in contrast to equilibrium cases where entropy difference between 
two states is measurable by heat produced in quasistatic processes.
Furthermore, even if we can measure nonequilibrium entropy, 
we cannot obtain unique temperature unless we properly set up 
the thermodynamic state space \cite{joubook, jou, muller, nettleton}; 
i.e. the value of the temperature depends on the choice of variables 
which we will fix through the differentiation of entropy 
with respect to energy \cite{joubook}.

Local equilibrium temperature, which we are familiar with, 
looses its validity for systems where the deviation 
from equilibrium ensemble is not negligible.
Indeed, it is expected that equipartition of energy will no longer be valid, 
in such a way that different degrees of freedom may have different energy.
For instance, some numerical simulations showed that nonequilibrium 
systems are anisotropic regarding with their kinetic energy \cite{baranyai}, 
which is never explained based on the local equilibrium assumption.

So far, several authors have tried to seek the meaning of temperature 
beyond the local equilibrium picture utilizing the microscopic expression 
devised by Rugh \cite{rugh}.
\begin{equation}
\label{rughtemperature}
\beta =\frac{\bra a\frac{N}{m}+b\nabla^2\phi\ket}
{\bra a\sum_i\frac{p_i^2}{m}+b|\nabla\phi|^2\ket},
\end{equation}
where $N$ is the number of degrees of freedom, $m$ denotes 
mass of the microscopic particles, 
and $\phi$ is the interparticle potential.
Note that we need arbitrary factors $a$ and $b$ 
to let the dimensions of the two terms 
(both of the numerator and of the denominator) be the same. 
We remark that, however, this arbitrariness has no influence 
on the value of temperature in equilibrium situations.
Although Eq. (\ref{rughtemperature}) is originally 
defined in microcanonical ensemble, 
Jepps et al. \cite{jepps} generalized this expression 
for canonical ensemble and presented it in a more general form.
Furthermore, they applied it to numerical simulations of 
nonequilibrium stationary states in the presence of shear flow or heat flow.
However, in nonequilibrium systems, we cannot have the unique value of 
temperature due to the arbitrariness of $a$ and $b$ \cite{ayton}.
This ambiguity seems quite natural since Rugh's expression is essentially 
based on equilibrium thermodynamic relation Eq. (\ref{def1}).
Namely, the problem is carried over from the choice of variables to be fixed: 
we cannot reach the dynamical expression of nonequilibrium temperature 
unless the correct thermodynamic state space is set up.

There is another way to define temperature 
which we call operational temperature in this paper. 
When finite closed systems are in contact,
they finally equilibrate to have the same intensive quantity, 
which we identify with temperature 
(i.e. the zeroth law of thermodynamics).
Hence, it might be possible to measure nonequilibrium temperature 
by putting equilibrium thermometer in contact to nonequilibrium systems.
A gedanken experiment has been proposed in order to 
clarify the meaning of nonequilibrium temperature 
from that point of view \cite{jou2}.
In particular, Baranyai has performed numerical simulations 
on shear-flow or heat-flow systems in contact with thermometers 
and obtained some explicit values of operational temperature 
\cite{baranyai}.
However, since we don't have a theoretical framework 
in which the obtained values should be interpreted, 
those numerical data seem to be left alone.
In other words, we cannot theoretically predict the value of 
operational temperature with which the numerical data should be compared.

Since the meaning of temperature out of equilibrium 
seems to lack a sound theoretical basis, 
it is reasonable to pick up the simplest model in order to analyze 
theoretical aspects more easily.
For that purpose, it seems that systems such as shear flow 
or heat flow are still excessively complicated.
In this paper, we adopt a forced harmonic oscillator as a model system.
Although it might be regarded as one of the simplest nonequilibrium 
systems, it is worth noting that time-averaged distribution functions 
of momentum and position are both Gaussian 
but characterized by different effective temperatures 
depending on the forcing frequency \cite{hatano}.
Thus, this system may provide a simpler example 
for the concept of temperatures than fluid systems under shear 
where different effective temperatures have been studied so far.
We will simulate an experiment analogous to \cite{jou2}, 
by letting a forced oscillator interact with another nonforced oscillator 
(in a different heat bath) which acts as a thermometer.
Comparison of the respective results may be useful 
for clarification of the concept of temperature.

The plan of the paper is as follows.
First, some statistical properties of a forced harmonic oscillator in 
a thermal bath are recalled and interpreted 
in terms of nonequilibrium temperature.
In the second section, in order to define temperature in 
a macroscopic point of view, a forced and an unforced oscillators 
situated in different thermal baths will be considered 
and the heat current between them will be calculated.
In the final section we discuss the form of the entropy for 
some different choices of variables, and compare our result 
with those obtained by Baranyai in the framework of 
nonequilibrium molecular dynamics of fluid systems of 
soft spheres in shear flow.

\section{A MODEL SYSTEM: FORCED HARMONIC OSCILLATOR}
We assume that our model system is described by 
the following Langevin equation;
\begin{equation}
\label{model}
\ddot{x}+\gamma\dot{x}+\Omega^2x=A\sin\omega t +\xi(t),
\end{equation}
where the mass of the oscillator is taken as unit.
The natural frequency of the harmonic oscillator is denoted by $\Omega$, 
and $A\sin\omega t$ corresponds to external forcing.
The noise term $\xi(t)$ is assumed to be Gaussian white noise 
which satisfies 
\begin{equation}
\label{noise}
\bra \xi(t)\ket =0, \ \ \ \ 
\bra\xi(t)\xi(t')\ket = 2\gamma\beta^{-1}\delta(t-t'),
\end{equation}
where $\beta$ is the inverse temperature of the heat bath.

Macroscopic or thermodynamic quantities should be defined by 
an appropriate averaging; usually ensemble-average or time-average.
Throughout this paper, we will take time-averaged quantities 
as macroscopic variables, since the model system is periodic 
in time due to sinusoidal forcing.

In our model, the external force gives power input to the oscillator, 
which may cause different influence on the average energy 
of momentum and of position. 
To have a physical idea of this influence, 
note that we have two kinds of relaxation times,
each of which is related to position and momentum, respectively.
We write the relaxation time of position as 
$\tau_x = \gamma\Omega^{-2}$ and the one of momentum as 
$\tau_p = \gamma^{-1}$.
When the forcing period $\tau_A=2\pi/\omega$ is longer enough 
than a relaxation time (i.e. $\tau_A\gg\tau_x$ or $\tau_p$), 
the sinusoidal motion is averaged to yield a distribution function 
which deviates from the equilibrium one.
In contrast, if the forcing period is comparable with (or shorter than) 
a relaxation time (i.e. $\tau_A\leq\tau_x$ or $\tau_p$), 
the corresponding motion of position or momentum cannot 
follow the forcing and the distribution function 
is indistinguishable from the equilibrium one.
For instance, if $\tau_p\ll\tau_A\leq\tau_x$, 
we may expect that the distribution function of position 
is not much changed from equilibrium, 
whereas the one of momentum is modified by the forcing.

To see these circumstances explicitly, we will calculate 
the potential energy $u\con$ and the kinetic energy $u\kin$.
Since the dynamics is linear, we can separate 
the ensemble-averaged motion and the fluctuation.
After some textbook-like calculations we get 
\begin{eqnarray}
\label{ucon}
u\con&=&\frac{1}{2\beta}+\frac{\Omega^2X_0^2}{4},\\
\label{ukin}
u\kin&=&\frac{1}{2\beta}+\frac{\omega^2X_0^2}{4}, 
\end{eqnarray}
where 
\begin{equation}
X_0= \frac{A}{\sqrt{(\Omega^2-\omega^2)^2+\gamma^2\omega^2}}.
\end{equation}
The corresponding energy dissipation rate $\dot{w}$ is 
\begin{eqnarray}
\dot{w}&=&\frac{A^2}{2}\frac{\gamma\omega^2}
{(\gamma\omega)^2+(\Omega^2-\omega^2)^2},\\
\label{dissipationrate}
&=&\gamma(2u\kin-\beta^{-1}).
\end{eqnarray}

Then we wish to consider the distribution function described by 
the Kramers equation corresponding to our model Eq. (\ref{model}).
\begin{equation}
\label{kramers}
\dot{\rho}(x,p;t)=\left[-\frac{\partial}{\partial x}p
+\frac{\partial}{\partial p}(\gamma p+\Omega^2 x-A\sin\omega t)
+\frac{\gamma}{\beta}\frac{\partial^2}{\partial p^2}\right]\rho(x,p;t).
\end{equation}
The solution independent of the initial condition is 
\begin{equation}
\label{ensemblepdf}
\rho(x,p;t)\propto\exp\left[-\frac{\beta}{2}
(p-\omega X_0\cos\omega t)^2
-\frac{\beta\Omega^2}{2}(x-X_0\sin\omega t)^2\right],
\end{equation}
where the time axis is shifted suitably.
Since this solution corresponds to ensemble distribution 
at a given instant, in order to calculate time-averaged quantities, 
the distribution function Eq. (\ref{ensemblepdf}) 
itself must be time-averaged,
\begin{eqnarray}
\nonumber
\rho(x,p)&\propto&\int_0^{\frac{2\pi}{\omega}}dt\rho(x,p;t),\\
\label{timepdf}
&\propto&\exp[-\frac{\beta}{2}(p^2-\Omega^2x^2)]
\int_0^{\frac{2\pi}{\omega}}dt
\exp[-\beta(p\omega X_0\cos\omega t-\Omega xX_0\sin\omega t)].
\end{eqnarray}
To perform this integral, we expand the integrand in Eq. (\ref{timepdf}) 
up to second order in $\beta p\omega X_0$ and $\beta\Omega^2xX_0$ 
(Gaussian approximation), providing that $\beta(\omega X_0)^2\ll 1$ 
and $\beta(\Omega^2X_0)^2\ll 1$.
Namely, we assume that gain of the internal energy due to 
the external forcing is small compared with thermal energy.
With this approximation, we get the following distribution function 
after simple calculations.
\begin{equation}
\label{gausspdf}
\rho(x,p)\propto\exp\left[-\beta\con\frac{\Omega^2x^2}{2}
-\beta\kin\frac{p^2}{2}\right], 
\end{equation}
where 
\begin{eqnarray}
\label{betacon}
\beta\con&=&\beta\left(1-\beta\frac{\Omega^2X_0^2}{2}\right),\\ 
\label{betakin}
\beta\kin&=&\beta\left(1-\beta\frac{\omega^2X_0^2}{2}\right).
\end{eqnarray}
We can see there are two kinds of temperature for one oscillator.
Hereafter each temperature corresponding to position and momentum 
is called configurational temperature and kinetic temperature, respectively.

\section{OPERATIONAL TEMPERATURE}
The discussion in the previous section deals with 
the distribution function in the phase space 
and hence it may be rather microscopic consideration.
From the thermodynamic point of view, the problem arises 
how the different microscopic temperatures of Eqs. (\ref{betacon}) 
and (\ref{betakin}) are connected to macroscopic measurements.
Since those temperatures differ, the macroscopically measured temperature 
(i.e. operational temperature) may indicate different values 
depending on the details of the connection 
between the system and the thermometer.
In this section, we will investigate the problem by devising 
a concrete model for the temperature measurement.

\begin{figure}
\scalebox{0.4}{\includegraphics{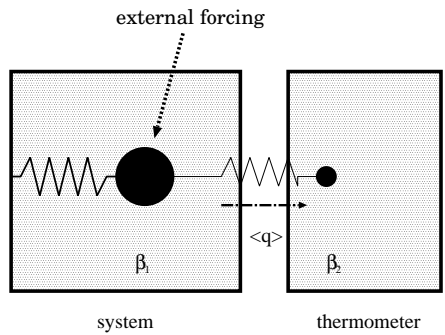}}
\caption{Schematic of the numerical experiment 
for operational temperature.}
\label{figure}
\end{figure}

Here we will examine a situation which bears some similarities with 
the proposal by Jou and Casas-V\'azquez \cite{jou2}.
They defined a prototype of operational temperature 
in which two systems are in thermal contact: 
the one is kept in a nonequilibrium steady state by means of heat flux, 
whereas the other is in equilibrium to act as a thermometer.
In our setting, we consider two coupled oscillators in contact with 
different heat baths whose temperature can be controlled independently.
The one is forced to stay away from equilibrium (the system) while 
the other is unforced to remain in equilibrium (the thermometer).
They are connected through a weak interaction potential $V$.
The schematic picture of our situation is shown in Fig. \ref{figure}.

The dynamics of such oscillators will be written as 
\begin{eqnarray}
\ddot{x}+\gamma\dot{x}+\Omega^2 x
+\epsilon\frac{\partial V(x-y)}{\partial x}
&=&A\sin(\omega t)+\xi_1(t),\\
\alpha\ddot{y}+\gamma\dot{y}
+\epsilon\frac{\partial V(x-y)}{\partial y}&=&\xi_2(t), 
\end{eqnarray}
where $\alpha$ denotes the mass of unforced oscillator.
We let $\alpha$ and $\epsilon$ to be small so that the disturbance 
of the forced oscillator by the thermometer will be weak.
The noise terms $\xi_i(t)$ are again assumed to be Gaussian white noise 
but characterized by different temperatures denoted by $\beta_i^{-1}$, 
i.e. 
\begin{equation}
\bra\xi_i(t)\ket=0, \ \ 
\bra\xi_i(t)\xi_j(t')\ket=2\gamma\beta_i^{-1}\delta_{ij}\delta(t-t'),
\end{equation}
where $\delta_{ij}$ is the Kronecker's delta 
(unity if $i=j$ and zero otherwise).
The similar system, in the absence of inertia and the external forcing, 
was studied in detail by Sekimoto \cite{sekimoto}.

We will test two kinds of interaction terms; 
harmonic and bistable potentials.
\begin{equation}
V(r)=\left\{
\begin{array}{@{\,}ll}
\frac{1}{2}r^2,\\
-\frac{1}{2}r^2+\frac{1}{4}r^4.
\end{array}
\right.
\end{equation}
The heat flux from the forced oscillator to the thermometer is 
evaluated as 
\begin{equation}
q=-\epsilon\frac{\partial V(x-y)}{\partial y}\dot{y}.
\end{equation}
When both oscillators are left unforced ($A=0$), 
the heat flux between both systems is proportional to the difference 
of the temperatures of the corresponding heat baths.
When one of the oscillators is forced ($A\neq 0$), 
the unforced oscillator plays the role of thermometer.

Our definition of the operational temperature is as follows.
We fix the parameters of the forced oscillator 
(i.e. $A$, $\omega$, $\beta_1$ and $\Omega$), 
and change the temperature of the heat bath for the thermometer ($\beta_2$).
There should be a certain value of $\beta_2$ at which 
the average heat flux $\bra q\ket$ vanishes.
Then we identify the temperature of heat bath $\beta_2^{-1}$ 
with the operational temperature.
Throughout the numerical simulations, we set $\gamma=1.0$, 
$A=1.0$, $\beta_1=1.0$, $\epsilon=0.1$, and $\alpha=0.1$.

\begin{figure}
\scalebox{0.4}{\includegraphics{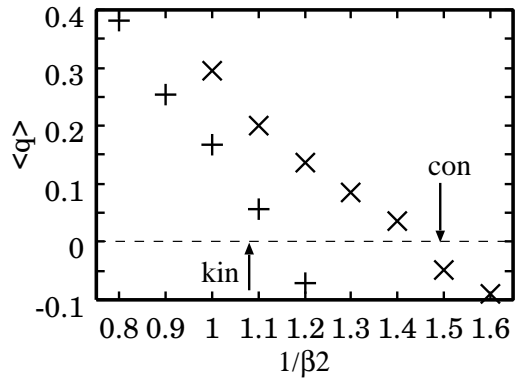}}
\caption{Average heat flux $\bra q\ket$ as the function of the temperature 
of the heat bath of thermometer.
The results of harmonic interaction are represented by $\times$'s, 
while the ones of bistable interaction are by +'s.
The parameters are set as $\omega=0.25$ and $\Omega=1.0$, 
where the corresponding kinetic and configurational 
temperatures are $1.08$ and $1.49$, respectively (shown by the arrows).}
\label{graph_con}
\end{figure}

\begin{figure}
\scalebox{0.4}{\includegraphics{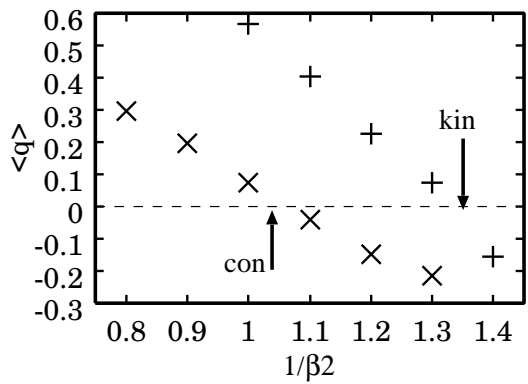}}
\caption{The same graph with Fig. \ref{graph_con} 
but with different parameters 
such that $\omega=1.0$ and $\Omega=0.1$, where 
the corresponding kinetic and configurational temperatures 
are $1.37$ and $1.04$, respectively (shown by the arrows).}
\label{graph_kin}
\end{figure}

The results of the numerical simulations are shown in 
Figs. \ref{graph_con} and \ref{graph_kin} for different parameters 
of the oscillator, where the values of $\bra q\ket$ are rescaled suitably.
Note that the zero-point of heat flux (equilibration point) 
is different depending on the interaction potential: 
thermometers indicate different values for the same system.
Especially, the one with harmonic potential shows 
good agreement with the configurational temperature, 
while the other with bistable potential indicates 
a value close to the kinetic temperature.
This tendency is unchanged for the system with different parameters 
at which the configurational temperature is higher than the kinetic one 
(Fig. \ref{graph_kin}), 
while the latter is higher than the former in Fig. \ref{graph_con}.

In short, the operational temperature of a thermometer 
with harmonic potential is almost the configurational temperature, 
and the one with bistable potential is close to the kinetic temperature.
It is concluded that various interactions show 
various temperatures whose values are ranged between kinetic and 
configurational temperatures.
Therefore, it is plausible that the zeroth law of thermodynamics 
should be formulated only in a very restrictive form 
for nonequilibrium cases.

\section{CONCLUDING REMARKS}
In this concluding section, we will pay some attention to entropy 
corresponding to the nonequilibrium system, and to a comparison 
with the results obtained by Baranyai and coworkers 
by means of the nonequilibrium molecular dynamics.

\subsection{Entropy for nonequilibrium steady states}

In this subsection we will see some possible 
extended Gibbs relations of forced harmonic oscillators.
The motivation is to discuss temperature in thermodynamic point of view 
and to compare it with the operational temperature 
obtained in the previous section.
The distribution function Eq. (\ref{gausspdf}) based on 
the Gaussian approximation may be considered to give 
the maximum entropy with the constraint 
that the second moments of momentum and position are given.
The corresponding Gibbs relation will thus be 
\begin{equation}
\label{gibbs1}
ds=\beta\con du\con+\beta\kin du\kin, 
\end{equation}
where $s$ denotes entropy per oscillator.
This may be a natural extension of the equilibrium Gibbs relation 
by considering that each degree of freedom is 
a thermodynamic system by itself with different temperature.
This simple form of the Gibbs relation is due to the Gaussian 
approximation of the distribution function. 
If they are not Gaussian, we would need additional 
independent variables; e.g. higher order moments.
Recall, anyway, that systems with two temperature are common 
in nonequilibrium physics, for example in plasmas 
(where electrons and ions may have different temperatures), 
or in metals or semiconductors (where electrons may exhibit 
a temperature different from that of the lattice).

However, Eq. (\ref{gibbs1}) is not the only candidate.
We can consider arbitrary linear transformation of $u\con$ and $u\kin$ 
such that 
\begin{eqnarray}
u&=&u\con+u\kin,\\
y&=&\alpha_1u\con+\alpha_2u\kin, 
\end{eqnarray}
where $\alpha_1\neq\alpha_2$.
By using these new variables, we can rewrite 
Eq. (\ref{gibbs1}) as 
\begin{equation}
\label{gibbs2}
ds=\frac{\alpha_2\beta\con-\alpha_1\beta\kin}{\alpha_2-\alpha_1}du
-\frac{\beta\con-\beta\kin}{\alpha_2-\alpha_1}dy.
\end{equation}
Since $u$ is the total energy of the oscillator, 
it may be possible to define temperature 
analogous to Eq. (\ref{def1}); 
\begin{equation}
\label{def2}
\theta=\left(\frac{\partial s}{\partial u}\right)_y, 
\end{equation}
which yields 
\begin{equation}
\theta=\frac{\alpha_2\beta\con-\alpha_1\beta\kin}{\alpha_2-\alpha_1}.
\end{equation}
Here we use another notation $\theta$ as 
the (inverse) temperature of the nonequilibrium system, 
which clearly depends on the choice of 
$\alpha_1$ and $\alpha_2$.
Namely, the thermodynamic temperature $\theta$ 
depends on the choice of the new variable $y$.

Although the nonuniqueness of temperature has been argued, e.g. 
in the context of extended irreversible thermodynamics \cite{joubook}, 
the criteria for the choice of new variables is still unknown.
At least there are some necessary conditions for the choice of $y$:
\begin{itemize}
\item it must be extensive, and macroscopically observable. 
\item the entropy must be convex regarding with the new extensive 
variable $y$.
\end{itemize}
In Eq. (\ref{gibbs1}), the convexity is identical with 
that $\beta\con$ and $\beta\kin$ are nonincreasing functions 
of $u\con$ and $u\kin$, respectively.
This is obvious since $\beta\con=(2u\con)^{-1}$ and 
$\beta\kin=(2u\kin)^{-1}$.
Of course there are other choices satisfying convexity.
Say we set $\alpha_1=1$ and $\alpha_2=0$, then 
\begin{equation}
\label{gibbs3}
ds=\beta\kin du+(\beta\con-\beta\kin)du\con, 
\end{equation}
where the thermodynamic temperature 
coincides with the kinetic temperature.

Another possibility is to adopt 
the entropy production rate $\sigma$ 
\begin{equation}
\sigma=\beta\dot{w}=\gamma(2\beta u\kin-1), 
\end{equation}
where Eq. (\ref{dissipationrate}) is recalled.
Since $d\sigma=2\gamma\beta du\kin$, 
from Eq. (\ref{gibbs1}) we get 
\begin{equation}
\label{gibbs4}
ds=\beta\con du-\frac{\beta\con-\beta\kin}{2\beta\gamma}d\sigma, 
\end{equation}
where the thermodynamic temperature becomes 
the configurational temperature.
Using Eqs. (\ref{betacon}) and (\ref{betakin}), 
we can further rewrite Eq. (\ref{gibbs4}) as 
\begin{equation}
ds=\beta\con du-\frac{1}{2\gamma^2}
\left(1-\frac{\Omega^2}{\omega^2}\right)\sigma d\sigma, 
\end{equation}
where we can see the second-order contribution of the flux 
$\sigma$ to the entropy.
Note that the above expression with energy and entropy production rate 
are analogous to that in extended irreversible thermodynamics 
\cite{joubook}, where the usual thermodynamic variables 
and the fluxes are taken as independent variables.
In this case we have $\beta\con$ as the thermodynamic temperature.
In addition, we remark that the second terms of 
Eqs. (\ref{gibbs3}) and (\ref{gibbs4}) vanish 
at resonance where $\omega=\Omega$. 
It is identical with 
\begin{equation}
\left(\frac{\partial S}{\partial y}\right)_u=0.
\end{equation}
Also the entropy production rate $\sigma$ is maximum at resonance.

As we have seen so far, thermodynamic temperature defined through 
extended Gibbs relation depends on the choice of the new variable.
We have expected that the operational temperature would be 
the criterion for choosing the new variable, 
which was one of the motivations of our study.
However, in the previous section, we have seen that 
different thermometers read different temperatures, 
which means impossibility of defining the unique nonequilibrium temperature 
even in this simple model system consisting of one degrees of freedom.
The absence of a unique operational temperature can be a serious problem 
for the construction of thermodynamics: 
at least the formulation of the zeroth law is not immediate 
and, if possible, would be a very restrictive form 
in contrast to equilibrium cases.
Sasa and Tasaki have already pointed out this kind of 
operational restriction which results from the anisotropy 
of pressure in a macroscopic heat conducting system \cite{sasa}.

\subsection{Comparison with the results of 
nonequilibrium molecular dynamics}
In this paper we have examined kinetic, configurational, 
and operational temperatures in a forced harmonic oscillator.
As was mentioned in the introduction, similar situations 
have been examined by Baranyai \cite{baranyai} based on 
the same motivation.
While he studied systems consisting of soft spheres under shear flow 
or in the presence of heat current using techniques of 
nonequilibrium molecular dynamics, 
the system analyzed here is much simpler than that.
It must be noted that Baranyai used the Nos\'e-Hoover type dynamics 
which removes energy from the system as dissipation, 
whereas we adopt the Langevin equation to represent the effect of heat baths.
Despite these differences, we think that it is still worth comparing 
these results since we are looking for general thermodynamic concepts 
which should be largely independent of the microscopic details of the system.

The definitions of temperature in the works of Baranyai is based on 
the Rugh's microscopic expression of Eq. (\ref{rughtemperature}).
As we have mentioned in the introduction, the expression itself cannot 
define temperature uniquely in nonequilibrium states due to arbitrary 
factors $a$ and $b$.
Instead of the unique temperature, again we have two kinds of temperature; 
configurational and kinetic temperatures which can be defined 
without the arbitrariness.
\begin{eqnarray}
\beta\con&=&\bra\frac{\nabla^2\phi}{|\nabla\phi|^2}\ket,\\
\beta\kin&=&\frac{\frac{N}{m}}{\bra\sum_i\frac{p_i^2}{m}\ket}.
\end{eqnarray}
Theses expressions yield $\beta\con=(2u\con)^{-1}$ and 
$\beta\kin=(2u\kin)^{-1}$ which coincide with our results 
obtained by Gaussian approximation.
Baranyai calculated these expression for temperatures in the mentioned 
fluid system of soft spheres at various values of the shear rate, 
and found that the configurational temperature is higher 
than the kinetic temperature, whereas the situation is opposite 
in systems with charge current.
Furthermore, these temperatures turned out to be anisotropic: 
i.e. they take different values for different spatial directions.
In our situation, the relation between the configurational 
and the kinetic temperatures depends on the ratio $\Omega/\omega$.
In general, as Baranyai has discussed, they will depend on 
the characteristics of the system and the external forcing 
responsible for the nonequilibrium situation.

In addition, Baranyai has proposed an operational temperature 
by devising the concrete model which emulates a physical thermometer 
in contact with the fluid.
The thermometer consists of the same particles as the fluid's, 
but do not feel the effect of shear flow nor the thermostatting: 
they interact only with the fluid particles.
This thermometer seems to read definite values of temperature 
regardless of the mass and the number of thermometer particles.
However, the dependency of interaction potential between the fluid and 
the thermometer is not discussed.
Taking our result into consideration, the operational temperature 
will depend on the interaction between the system and thermometers.
Indeed, Hoover et al. have discussed the ideal gas thermometer 
which reads the kinetic temperature \cite{hoover}, while Baranyai's 
thermometer reads the value which is closer to the configurational 
temperature.
However, since Hoover et al. ignored the anisotropy of the kinetic energy, 
it is not apparent what value the ideal gas thermometer reads 
when it is actually in contact with a nonequilibrium system.
More numerical simulations and real experiments on the operational 
temperature are needed for the clarification of nonequilibrium 
temperature and the underlying nonequilibrium thermodynamics.

In summary, we have found that the operational temperature depends on 
the details of interaction between the system and a thermometer, 
which is never seen in equilibrium situations.
The fact may shove a strong restriction on the extension of 
the zeroth law for nonequilibrium systems.
Of course, as our analysis is confined to one-dimensional systems, 
there may arise another problem for two- or three- dimensional 
systems regarding with the relation between the anisotropy 
of temperatures and the operational temperature.
In addition, the Gaussian approximation of the distribution 
functions make especially easy to define temperatures, 
while additional conceptual problems would appear 
if the distribution function deviates far from the Gaussian.
For instance, there have been some maximum-entropy analyses of 
nonequilibrium radiation, where nonequilibrium temperature or 
quasitemperature per mode have been defined in terms of 
the nonequilibrium populations of the different modes, 
in the context of a generalized Planck statistics 
instead of a classical Boltzmann statistics \cite{vasconcellos}.
The present situation has the advantages of the higher simplicity and 
of the possibility of devising numerical simulations concerning 
operational temperature which have not been done before.

\acknowledgments
D. Jou acknowledges the financial support of the Spanish 
Ministry for Science and Technology under grant 
BFM2000-0351-C03-01 and of the General Direction for 
Research of the Generalitat of Catalonia under grant 2001 SGR 00186.
T. Hatano acknowledges the financial support of 
Japan Society for Promotion of Science under grant 12-9288.

\end{document}